\documentclass[pdf, 10pt, twocolumn]{article}
\usepackage{authblk}

\usepackage{xcolor}
\usepackage{amsmath}
\usepackage{amssymb}
\usepackage{lineno}
\usepackage{array}
\usepackage{graphicx}
\usepackage{microtype}

\begin{document}

\title{A Neutron Microscope Using a Nested Wolter-I Condenser and a Bank of Diffractive–Refractive Achromatic Objectives}

\author[1]{\small H.F. Poulsen}
\author[2]{\small C.L. Andersen}
\author[2]{\small N. Ravinet}
\author[3]{\small J. Vila-Comamala}
\author[3]{\small M.R.D. Veeraraj}
\author[1]{\small E.B. Knudsen}
\author[1,4]{\small P.K. Willendrup}
\author[2]{\small S. Massahi}
\author[2]{\small F.E. Christensen}
\author[5]{\small D.D.M. Ferreira}
\author[6]{\small M. Strobl}
\author[3]{\small C. David}
\author[7]{\small L. Theil Kuhn}


\date{\today}

\affil[1]{\small Department of Physics, Technical University of Denmark, 2800 Kgs. Lyngby, Denmark}
\affil[2]{\small CHEXS ApS, Diplomvej 373B, 2800 Kgs. Lyngby, Denmark}
\affil[3]{\small PSI Center for Photon Science, Paul Scherrer Institute, Forschungsstrasse 111 5232 Villigen, Switzerland}
\affil[4]{\small European Spallation Source ERIC - DMSC, 2800 Kgs.~Lyngby, Denmark}
\affil[5]{\small Department of Space, Technical University of Denmark, Elektrovej 327, 2800 Kgs. Lyngby, Denmark}
\affil[6]{\small PSI Center for Neutron and Muon Sciences, Paul Scherrer Institute, Forschungsstrasse 111 5232 Villigen, Switzerland}
\affil[7]{\small Department of Energy Conversion and Storage, Technical University of Denmark, Fysikvej 310, 2800 Kgs. Lyngby, Denmark}


\maketitle                        


\begin{abstract}

We propose a nested Wolter-I mirror design for a neutron condenser, which is based on established X-ray telescope technology. We demonstrate through simulations that it can increase the flux density at the ESS imaging instrument ODIN by up to two orders of magnitude. Experimental measurements of reflectivity and figure errors on a prototype mirror element confirm the technical feasibility of the approach. Then, we discuss design strategies for an imaging objective to fully exploit the condenser specifications while achieving spatial resolutions comparable to those of X-ray micro-CT instruments. Analytically, we show that for monochromatic beams suitable solutions exist employing arrays of hundreds of identical objectives, realized either as compound refractive lenses (CRLs) or Fresnel zone plates (FZPs). To mitigate the inherent chromatic aberration of these optics, each individual objective could be replaced by an achromatic FZP/CRL combination. Key optical properties of the resulting microscope are estimated. This novel full-field microscopy concept for highly divergent, polychromatic neutron beams has the potential to improve temporal and spatial resolution for large samples and sample environments and to enable the simultaneous acquisition of hundreds of projections in neutron tomography.\\
\end{abstract}

\vspace{1.2cm}

\section{Introduction}

Neutron imaging is an expanding field \cite{Lehmann2015, Kardjilov2018, Treimer2019, Strobl2024}. The relatively low brilliance of neutron sources implies that imaging experiments typically require the use of a polychromatic beam with a relatively large divergence — in particular this is true for 3D and time resolved studies. With such beams, it is natural to base bright field imaging studies on placing a 2D detector downstream in close proximity to the sample. This pinhole camera approach enables a large variety of contrast to be explored \cite{Strobl2024}, including attenuation contrast \cite{Kallmann1948, Strobl2009}, phase contrast \cite{Allman2000, Paganin2023, Oestergaard2023}, dark field imaging \cite{Pfeiffer2006b, Strobl2008b, Bacak2020, Shen2025}, 
diffraction contrast imaging \cite{Santisteban2001, Lehmann2014}, grain mapping  \cite{Peetermans2014, Cereser2017, Larsen2025}, 
and some versions of polarized neutron imaging for visualising magnetic field distributions \cite{Kardjilov2008,Sales2018, Sales2019, Strobl2019}, magnetic domains \cite{Manke2010, Hiroi2018, Jorba2019} and current distributions \cite{Karimi2025}. 

One basic limitation, however, is that the resolution deteriorates with larger distances between the sample and the detector, unless one compromises on the divergence of the incoming beam. This implies a trade-off between spatial resolution and time resolution, and, in general, limits the spatial resolution for extended objects and extended sample auxiliaries as well as for modalities such as polarized imaging where extended optics are required.

Inspired by imaging with light, X-rays and electrons, it is natural in such cases to consider a \textbf{neutron microscope} - comprising a condenser optic upstream of the sample and an objective lens between the sample and the detector. The magnification intrinsic to most microscopes may also allow the use of more efficient detectors. Alas, such an instrument is yet to be realised.

In practice, several types of neutron optics have been demonstrated. Refractive optics, Compound Refractive Lenses (CRLs), may provide high spatial resolution \cite{Eskildsen1998, Beguiristain2002, Cremer2005}. However, the numerical aperture NA is fundamentally limited by the critical angle  $\alpha = \sqrt{2\delta}$, where $\delta$ is the refractive index decrement. The resulting NA at 4 \AA{} is at best 0.01~rad. Likewise, diffractive optics elements, Fresnel Zone Plates, FZPs, have been presented  \cite{Kearney1980, Sacchetti2004, Veeraraj2025}, but here the NA is limited by the manufacturing process to a few mrad. In comparison, the divergence accepted by the primary optics (in-pile neutron guide) is typically of the order of a few degrees ($\sim 0.03$ rad), implying that refractive and diffractive optics introduced downstream would only transmit a fraction of the incoming beam. Moreover, both types of optics are chromatic. 

As an alternative, it has been suggested to use nested concentric reflecting mirrors \cite{Mildner2011, Liu2013, Abir2020}, either in a Wolter-I geometry (a paraboloid followed by a hyperboloid mirror) or a Wolter-II design (a convex paraboloid mirror followed by a concave hyperboloid mirror). This type of optics is well matched to the large divergence and size of the neutron beam and can be made achromatic using super-mirrors.  As a result, the potential increase in neutron flux is several orders of magnitude. However, the requirements for the shape errors of mirrors are very demanding: of the order 1~mrad for a condenser with a focal length of 1~m, and 1-10 $\mu$rad for an objective of focal length of 1~m. 

In this paper we present progress towards realizing a neutron microscope. The presentation will focus on high flux reactor and spallation sources and more specifically on an optic for full field neutron imaging at the European Spallation Source, ESS. 

Initially, we present a design for a Wolter-I type condenser, that mimics the design of an existing X-ray telescope.  The neutron condenser has a focal length of 1~m, and comprises 20 shells  with diameters ranging from 1 to 5~cm. To reflect neutrons effectively over a broad wavelength range the individual depth-graded mirrors comprise multilayer coatings of Ni/Ti or NiC/Ti. We highlight the potential by showing full scale ray tracing simulations by means of the McStas program \cite{Willendrup2020},\cite{Willendrup2021} for a test at the upcoming ODIN beamline at ESS \cite{Strobl2015}. Despite the fact that this prototype condenser is not optimized for the beamline we find it may provide a flux density gain of 100 times in a focal spot size of 4 $\times$ 2 mm$^2$. Next, we present the results of feasibility tests for a single 60-degree mirror element of this design.

The resulting condensed beam can only be used for pinhole type imaging of thin samples, as the divergence of the beam incident on the sample has increased from 0.5~deg to 3.0~deg, and consequently the depth of field has deteriorated. Moreover, performing tomography with such a beam implies that each projection is an average over a 3 degree rotation, which often is unacceptable. A spot size of 4 $\times$ 2 mm$^2$ is also in general too big for scanning modalities. However, a highly intense 4 $\times$ 2 mm$^2$ beam provides an adequate field-of-view for microscopy, provided a suitable objective lens can be realized. As already mentioned, using a nested mirror design for the objective is 2-3 orders of magnitude more demanding in terms of shape errors. Instead, we propose a novel concept for the objective design based on combining diffractive and refractive optics elements. A spatial resolution of 2-10 $\mu$m is inherent to such optics, the challenge is how to make them compatible with the large divergence in the condensed beam and relevant for a broad energy band. 

The solution suggested is \textbf{a bank/array of objectives, each of them being a neutron achromat}. In the second part of the paper, we review state-of-art of neutron CRL and FZP technology and achromat design based on combining a converging FZP and a diverging CRL  \cite{Poulsen2014}.  We describe the NA, field-of-view and bandwidth of banks of such achromats. Finally, we discuss the technical feasibility and outline the science case for an instrument combining the condenser and the objective bank. Here one overriding ambition is to enable 3D neutron imaging with the same spatial and angular resolution as state-of-the-art X-ray micro-CT, thereby addressing the needs of a large community in materials science.


\section{Design and feasibility tests of Wolter-I type condenser}
\label{sec-condenser}

We have chosen a design that is an adaptation of the technology in the  NuSTAR-telescope probing 3-79~keV X-rays \cite{Harrison2013}. The slope errors for such telescopes can be below 1~arc minute, which is sufficient for a neutron condenser.  The specific multilayer coating consists of a depth-graded multilayer following a power-law function \cite{christensen2011a}.

\subsection{Ray tracing}\label{s:rt}

For ray tracing we have applied an idealized $m=3$ super-mirror model coating based on Ni and Ti layers. The selected coating model does not include figure error but includes a phenomenological model roughness for planar $m=3$ mirrors, derived from Swiss Neutronics experimental data \cite{Jacobsen2013}.

\begin{figure}
    \centering
    \includegraphics[width=1\columnwidth]{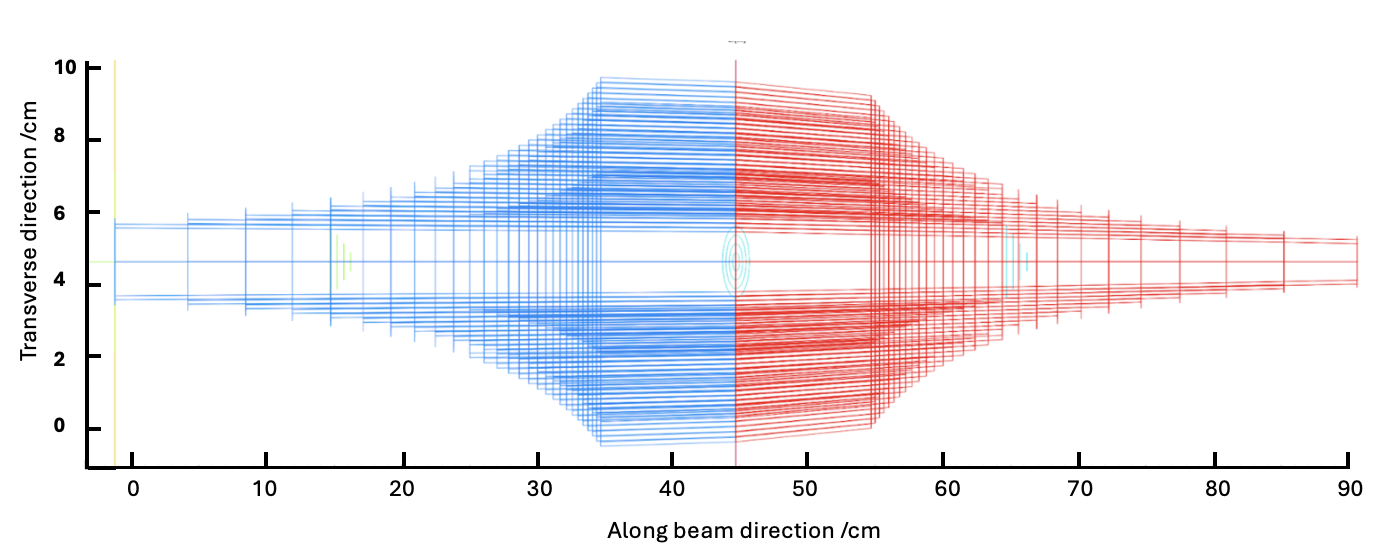}
    \caption{\small Wolter optic based condenser design. Line rendering of the proposed focusing optic with 20 shells of conical symmetry.}
   \label{fig:condenser_design}
\end{figure}


The design is outlined in Fig.~\ref{fig:condenser_design}. A total of 20 concentric conical layers are applied, all with a primary and secondary mirror in a Wolter-I geometry.  The radii vary from 1 to 5~cm. To achieve full coverage of this range, the mirrors have been designed with variable length, such that the inner mirrors are significantly longer than the outer ones. The nominal focal length of the optics, as measured from the center line of the optics is 1~m. Hence, the set-up allows for 55~cm between the exit of the optics and the pivot point of the sample goniometer. 

The simulations were performed using McStas \cite{Willendrup2020},\cite{Willendrup2021}, specifically targeting a potential test of the optics at ODIN:  an upcoming  time-of-flight instrument with a peak wavelength at 3.7 {\AA} \cite{Andersen2020}. For the simulations, a model from the design phase of ODIN \cite{Strobl2015, Schmakat2020} was used. The model includes a set of straight neutron guide sections with rectangular cross-section and elliptical shape along the horizontal and vertical planes. The final elliptical section ends with a rectangular aperture of size $5\times 2.5~$cm$^2$ (W $\times$ H). The McStas implementation follows the concepts developed in the AstroX-project \cite{Knudsen2018}, where the mirrors are considered to have fully opaque back-walls, \emph{i.e.} there is no cross talk between the mirrors. 
\begin{figure*}
    \centering
    \includegraphics[width=0.7 \textwidth]{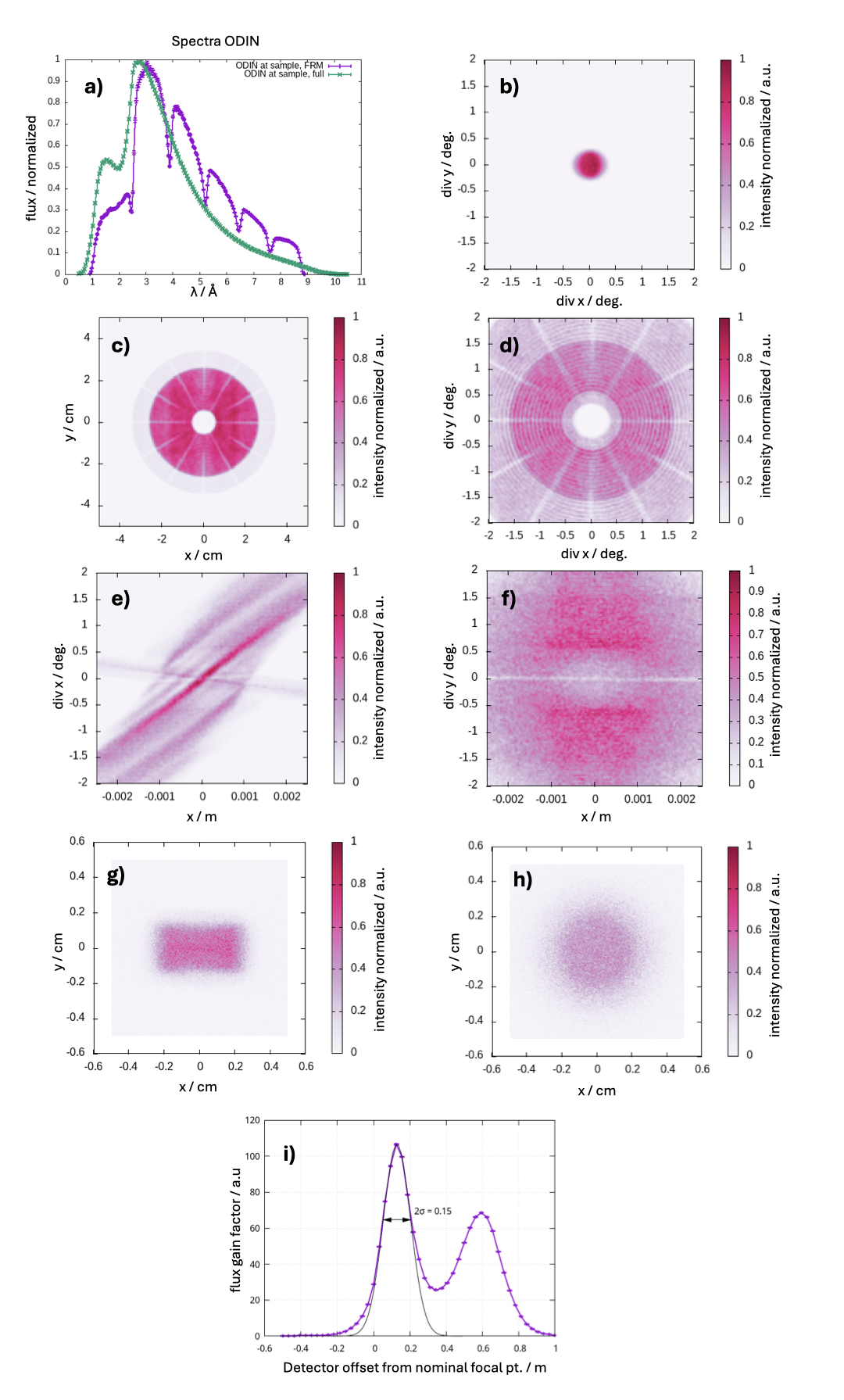}\caption{\small Results of McStas simulations of the Wolter-I type condenser for use at the ODIN instrument. a) Spectrum of ODIN with (purple) and without (green) the Frame Rate Multiplication, FRM, chopper activated. b) The divergence of the beam just upstream of the condenser. c) spatial distribution of the beam just downstream of the condenser. d) 2D-divergence distribution, e) and f) horizontal and vertical divergence as function of position along the beam, x, g) spatial distribution, all of d)-g) measured at the focal point. i) Gain factor of neutron flux as function of distance from the nominal focal point. The first peak in i) corresponds to the focal point, the second peak stems from a geometrical artifact caused by a single reflection on the Wolter-I optic. The resulting artificial beam spot at 0.6~m is correspondingly diffuse, see h). 
    }
   \label{fig:knudsen}
\end{figure*}
In Fig \ref{fig:knudsen}a) the wavelength spectrum at the exit pinhole of the neutron guide is shown. Gaussian fits to the divergence distribution at this pinhole result in widths (FWHM) of 1.03~deg and 1.06~deg in the horizontal and vertical directions, respectively. 
For the existing design of ODIN, the best position of the optics will be 10~m downstream of the pinhole. The dimensions of our Wolter-I optics design accommodate this.  In Fig \ref{fig:knudsen}b) the divergence of the beam at the \emph{entry} of the condenser is shown. We determine the angular spread of this beam to be 0.50~deg. 
In Fig \ref{fig:knudsen} c) the intensity distribution just after the condenser is shown - the spokes keeping the layers apart are clearly visible, as well as the central "20 mm hole" which is not mirrored (this may either be blocked or  used for a low-divergent modality in parallel). 

In Fig. \ref{fig:knudsen} d) the angular distribution at the focal point is shown. This is nearly uniform (except for the central hole and the spokes) with a FWHM of approx. 3.0~deg (50~mrad). In the actual focal spot at a distance of 1.15 m from the center of the condenser - Fig. \ref{fig:knudsen} g) - the spot size is rectangular with a width of approx. 4 mm $\times$ 2 mm. Finally in Fig. \ref{fig:knudsen} i) the gain in neutron flux in relation to the unfocused case is shown as function of distance from the nominal focal point. The "focal spot" is seen to be elongated over 15~cm along the beam and exhibit a gain factor of 100 at the optimal distance. The secondary peak around 0.6 m is an artifact of stray neutrons, singly reflected through the optic.

\subsection{Feasibility tests}

To assess the practical feasibility of a Wolter-I type condenser, supermirror-coated, curved glass substrates were fabricated at the company CHEXS. We formed thin glass into 60-degree cylindrical segments with a curvature radius of 20~mm and a length of 100 mm. The glass is coated with a 256-bilayer NiC/Ti supermirror designed by CHEXS, utilizing DC magnetron sputtering. This coating targets an m-value of 3, corresponding to a minimum bilayer thickness of 9.6~nm. For a complete shell in the Wolter-I optics design, six such segments are required.

\begin{figure}
    \centering
    \includegraphics[width=0.9\columnwidth]{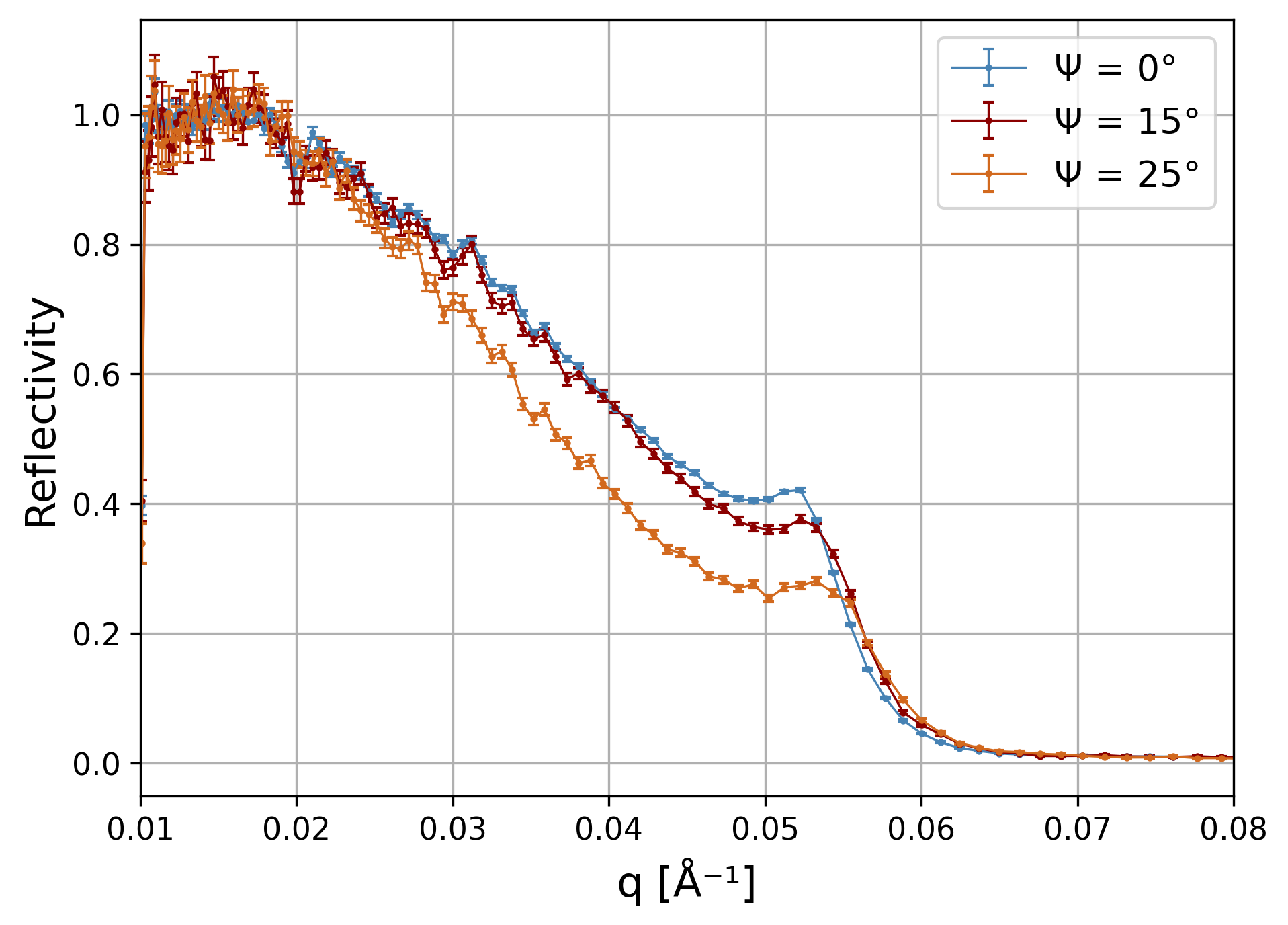}
    \caption{\small Measured neutron reflectometry of curved NiC/Ti supermirror-coated glass at three azimuthal positions, $\Psi$.}
   \label{S020_data}
\end{figure}

A supermirror segment was measured at the neutron reflectometer AMOR at PSI. Specular neutron reflectivity was recorded at three azimuthal positions 
($\psi = 0^\circ$, $15^\circ$, and $25^\circ$), with $\psi = 0^\circ$ set at the center 
without distinction between the two symmetric sides of the arc. The resulting data are shown in Figure~\ref{S020_data}. The measurements show that the mirror segment achieved an m-value of 2.5, which corresponds to a minimum bilayer thickness of 11.2~nm. This coating parameter will be further optimized in subsequent experiments. The mean reflectivity at the three positions was found to be 0.70, 0.67, and 0.60, decreasing from the center towards the edges. 

An uncoated substrate was X-ray CT scanned, and the preliminary  results indicate a figure error better than 1~arc minute. Additional characterization measurements are currently being performed on the coated substrates. 

These results demonstrate that our first iteration of supermirror-coated curved glass substrates show promising performance and support further development towards a Wolter-I neutron condenser with performance comparable to that of the McStas simulations. 

\section{Objective bank solutions for a monochromatic beam}
\label{sec-CRLobjective}

In this section we survey the possibilities for designing a bank (a 2D array) of either refractive or diffractive objectives subject to the following conditions
\begin{itemize}
    \item  A field of view of approximately 2 $\times$ 4 mm$^2$ corresponding to the spot size provided by the condenser described in section \ref{sec-condenser}.
    \item  An angular acceptance corresponding to the divergence of the incident beam, that is 3~deg (0.052~rad).
    \item A spatial resolution corresponding to state-of-the-art micro-CT machines: in the range 2 - 10 $\mu$m. 
\end{itemize}
As both refractive and diffractive elements are chromatic we shall assume a monochromatic beam.

To facilitate the presentation, for both types of optics we initially summarize key expressions for a single optical element.

\subsection{Geometry of a single neutron CRL objective}

The properties of  a single neutron CRL objective can be expressed analytically using geometrical optics. Following  \cite{Simons2017} the CRL is assumed to comprise $N$ identical lenslets,  characterized by their radius of curvature at apex, $R$, and the distance between the centers of neighboring lenslets, $T$. The focal length for each lenslet is then $f = R/(2\delta)$, while the  physical aperture is $2Y = \sqrt{RT}$. The focal length for the entire CRL, $f_{CRL}$, is 
\begin{eqnarray}
f_{CRL}  = & f \varphi \cot(N\varphi) = &\sqrt{Tf}\cot \Big( \frac{NT}{\sqrt{Tf}}\Big). \label{eq-fN}
\end{eqnarray}
Here the shorthand $\varphi = \sqrt{T/f}$ is introduced. 

Next, we consider an imaging system with $d_1$ and $d_2$ being the distances from the object plane to the entry of the CRL, and from the exit of the CRL to the image plane, respectively.
For the (unsigned) magnification we have 
\begin{eqnarray}
 \mathcal{M}  = & \dfrac{1}{\cos(N\phi)} \dfrac{f_{CRL}}{d_1-f_{CRL}}  =  \cos(N\phi) \dfrac{d_2-f_{CRL}}{f_{CRL}}.  \label{eq-M}
\end{eqnarray}

Closed expressions for the numerical aperture, $NA$, and the field-of-view, FOV,  are derived in \cite{Simons2017}. Following conventions in that work CRL numbers for FOV and NA numbers in this paper represent the FWHM. In \cite{Leemreize2019} it is argued that a neutron CRL typically can be approximated by a \emph{transparent lens}, where the attenuation of the neutrons within the parabolic part is neglected.  For some optical properties the CRL then simply acts as a collimator with dimensions given by the physical aperture $2Y$ and the length $NT$. The numerical aperture is in this case
\begin{eqnarray}
NA  = \min \Big( \frac{2Y}{d_1}\cos ( N\varphi), 2\sqrt{2\delta}\sin(N\varphi) \Big). 
\label{eq-NA}
\end{eqnarray}
The first term reflects the limitation by the CRL as a collimator, the second term its limitation in terms of refractive power. The cosine factor in the first term originates in the fact that the neutron trajectory within the CRL approximately is a sinusoidal with period $2\pi N\varphi$.  In the transparent lens case we have the following expression for the field-of-view
\begin{eqnarray}
FOV  = 2Y \cos(N\varphi)\Big(\frac{d_1}{NT}+1 \Big). 
\label{eq-FOV}
\end{eqnarray}

The maximum NA available is realised for $N\varphi = \pi/4$. Then  Eq. \ref{eq-fN} and Eq. \ref{eq-FOV} become
\begin{eqnarray}
     f_{CRL} &=   2Y/(4\sqrt{\delta}) = \frac{\pi}{4\sqrt{2}} \frac{f}{N}, 
     \label{eq-fN_v2} \\
     FOV  & = \sqrt{2} Y (\frac{d_1}{f_{CRL}}+1) \approx \sqrt{2}\;  2Y,
\label{eq-FOV_v2}
\end{eqnarray}
where the approximation is valid for large $\mathcal{M}$.

\subsection{A CRL based objective bank}
\label{sub-bankCRL}

Following \cite{Leemreize2019}, we select two materials as prime candidate materials for scale up: diamond, due to its superior specifications and prior use, and MgF$_2$, which is easily available and can be machined by milling machines.  Moreover, both are readily available as single crystals. This is important as both materials exhibit a pure absorption cross section that is two orders of magnitude smaller than the coherent scattering cross section.

In Table~\ref{tab:opticals_constants}, we provide key numbers. The tabulated values are for cold neutrons, as these are relevant for most of the applications presented. Writing the refractive index $n = 1 - \delta$, the decrement $\delta$ is positive for both materials - similar to the case for X-rays - but an order of magnitude larger. We have $\delta \propto \lambda^2$, where $\lambda$ is the neutron wavelength.

\begin{table*}
    \centering
    \begin{tabular}[width=1\columnwidth]{|c|cccc|}
         Material &  $\rho$ (g/cm$^3$)  & $\delta (10^{-6})$ &
         NA$_{max} (10^{-3})$ & $\mu$ (m$^{-1}$) \\
         C (Diamond) &  3.51 & 29.3  & 10.8    & 0.08  \\
         MgF$_2$ & 3.15  & 7.1  & 5.02  & 0.44  \\
    \end{tabular}
    \caption{\small Optical properties of materials for neutron CRLs: $\rho$ is density, $\delta$ the refractive index decrement and NA$_{max} = \sqrt{2\delta}$ the maximum Numerical Aperture. $\mu$ is the linear attenuation coefficient for a single crystal that exhibits no diffraction. The numbers relate to a wavelength of 4 \AA.} 
    \label{tab:opticals_constants}
\end{table*}

\begin{figure}
    \centering
    \includegraphics[width=1\columnwidth]{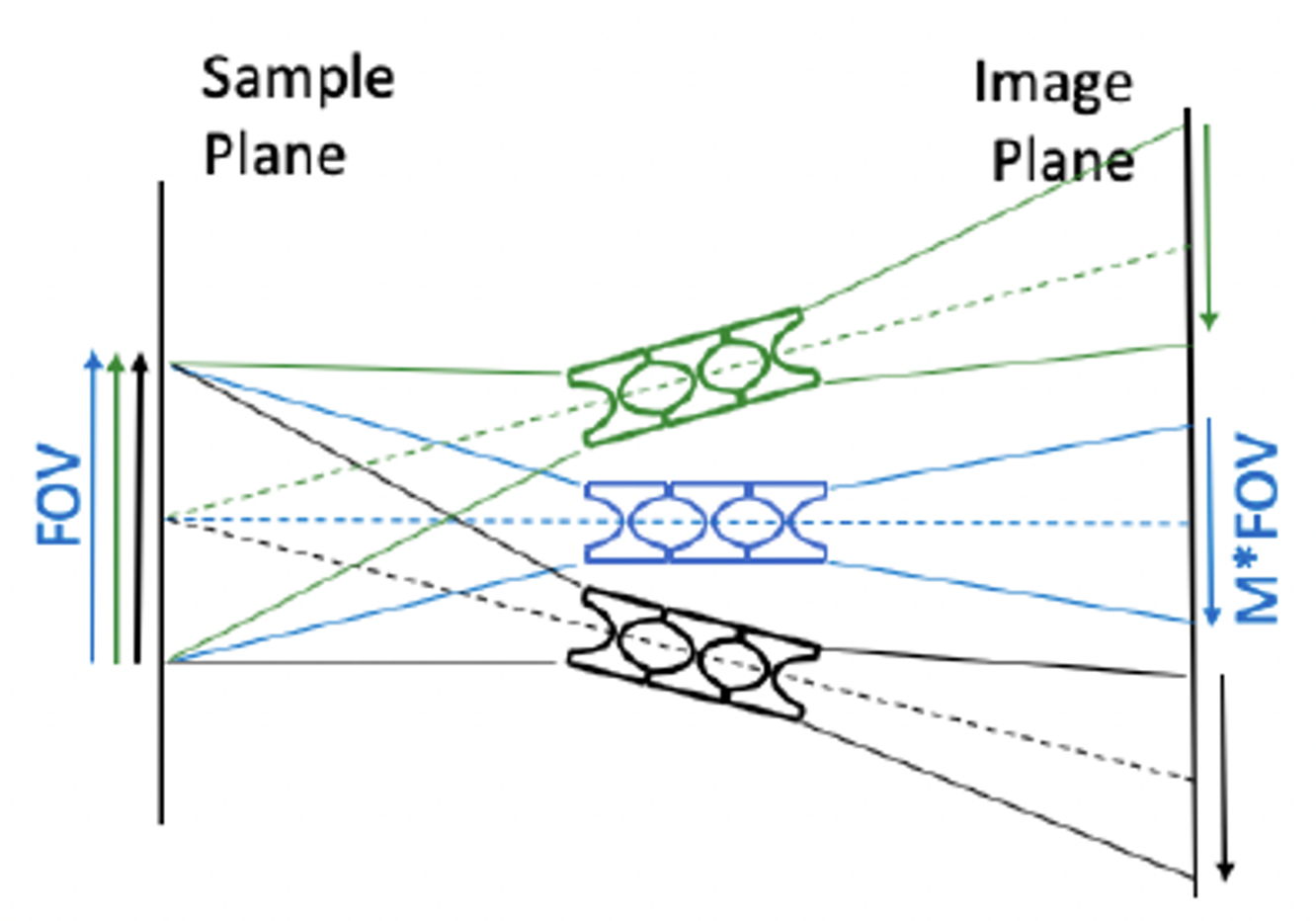}
    \caption{\small Illustration of the concept of a bank of objectives, where each objective is a compound refractive lens. The optical axes (dashed lines) coincides in the sample plane. Adapted from \cite{Leemreize2019}.}
   \label{fig:CRL_objective__bank}
\end{figure}

It appears that the incident divergence provided by the Wolter-I condenser is about 5 times larger than the maximum NA for a C-based CRL objective, and about a factor of 10 times larger in the case of a MgF$_2$-based CRL objective.  Central to this work is the suggestion to overcome this hurdle by using a 2D bank of objectives, tilted such that the optical axes of the various CRLs coincide in the Origo of the sample plane. This is illustrated in one plane in
 Fig. \ref{fig:CRL_objective__bank}.

There are four independent variables: $N,R,T, d_1$ that determine the four optical properties $f_{CRL}, \mathcal{M}, NA$ and  the distance $d_2$. Reversely, specifying the optical properties and/or the length of the laboratory (and thereby $d_2$) defines a design of the CRL and the working distance $d_1$.  We now discuss optimizing these parameters to conform as closely as possible to the condenser specifications.  

First we aim for the maximum achievable NA, tabulated in Table~\ref{tab:opticals_constants}, as this will reduce the number of CRLs.  Requiring $FOV = 4$ mm implies that $2Y = 4/\sqrt{2}$ mm. From Eq.~\ref{eq-fN_v2} follows that the focal length $f_{CRL} = 130$ mm for diamond and $f_{CRL} = 265$ mm for MgF$_2$. As an example for C this is achievable by the design $(N,R,T)$ = (59, 1 mm, 2.8 mm), which is technically feasible. Notably, the maximum NA appears in the thick lens limit, here $NT = 46$ cm. 

In terms of resources such a set-up would require approximately 25 (C) or 100 (MgF$_2$) individual CRLs and an appropriate detector coverage. When relevant, the intensities in the individual images may be superposed to form one resulting 2D image of the 4 $\times$ 2 mm$^2$ region of interest. For tomography, each  column of the array provides one projection in the neutron tomogram. 

Notably, an implementation of a bank of CRLs for the analogous case of a Multi-Lens Array Full-Field X-ray Microscope (for X-ray laboratory sources) is presented in \cite{Opolka2021}. Here manufacturing is done by means of lithography.

An additional constraint applies to ensure that the partial images provided by neighboring CRLs do not overlap on the detector. The condition for a complete spatial and angular sampling is  
\begin{eqnarray}
\mathcal{M} \le \frac{NA (d_1+NT+d_2) }{FOV} \label{Eq-sampling}
\end{eqnarray}
In practice this condition as well as the dimensions of the instrument restricts the magnification. Still magnification by 5-10 is feasible, which may open up for imaging with modern day pixel detectors.

In conclusion, for monochromatic beam operation, this objective bank is well suited for inclusion in a micro-CT neutron microscope based on the Wolter-I condenser. The available space around the sample is adequate, it matches the field-of-view and the magnification enables the combination of a micrometer-sized spatial resolution with a magnification of 5 or more. Moreover, the depth of field DOF = $y_s/NA$ (with $y_s$ being the spatial resolution in the sample plane) will be hundreds of micrometers, which is acceptable for many science cases.

\subsection{Geometry of a single FZP based objective}
\label{sec-FZPobjective}

A Fresnel zone plate (FZP) is a diffractive optical element composed of concentric rings with radially decreasing line widths.  In the following we consider a neutron phase-FZP where the  rings/zones alternate between being phase-shifting and non–phase-shifting \cite{Kearney1980, Sacchetti2004, Altissimo2004, Veeraraj2025}. Phase FZPs have a higher diffraction efficiency than amplitude FZPs and they also suppress the zeroth order diffraction beam, making them better suited for imaging applications. 

An ideal phase Fresnel zone plate has a maximum diffraction efficiency of 40.5 \% that can be achieved by choosing a zone thickness $\Delta T_{\pi}$ that induces a $\pi$ phase shift in the incident radiation, 
\begin{equation}
    \Delta T_{\pi} = \lambda/(2\delta). \label{Eq-FZP-T_pi}
\end{equation}
For thermal and cold neutrons, with wavelengths ranging from 1 to 10 \AA{}, $\Delta T_{\pi}$ required is typically on the order of micrometers to a few tens of micrometers. The diffraction efficiency can be significantly improved by using blazed FZP, in which the zone profiles consist of steps to approximate a continuous phase ramp \cite{DiFabrizio1999, Mohacsi16}.

The optical properties of the FZP are governed
by the outermost zone width, $\Delta r$.
For thermal or cold neutrons the focal length of a single FZP, $f_{FZP}$, will be much larger than the diameter of the optic, $D$. In the paraxial approximation  we have
\begin{eqnarray}
    f_{FZP} = \frac{D \, \Delta r}{\lambda}, \label{FZP-f} \\
    NA = \frac{D}{2f_{FZP}} = \frac{\lambda}{2\Delta r}. \label{FZP-NA}
\end{eqnarray}
The FZP being a thin lens implies the following relationships in an imaging set-up
\begin{equation}
  \begin{aligned}
    \frac{1}{f_{FZP}} = \frac{1}{d_1} + \frac{1}{d_2};
    \mathcal{M}  = \frac{d_2}{d_1}; \\
    d_1 = \frac{L}{\mathcal{M}+1}; 
    d_2 = \frac{\mathcal{M} \; L}{\mathcal{M}+1},
  \end{aligned}
\end{equation}
with $L$ being the length of the imaging system: $L = d_1 + d_2$. 

The Field-of-View of an FZP is discussed in e.g. \cite{Howells2017}. While aberrations exist for off-axis imaging, for e.g. 5 keV X-rays these remain below 100~nm for objects with a size as large as 10 times the zone plate radius. From this we assume that such distortions will not have a noticeable impact on the resolution in neutron microscopy.

\begin{figure}
    \centering
    \includegraphics[width=1\columnwidth]{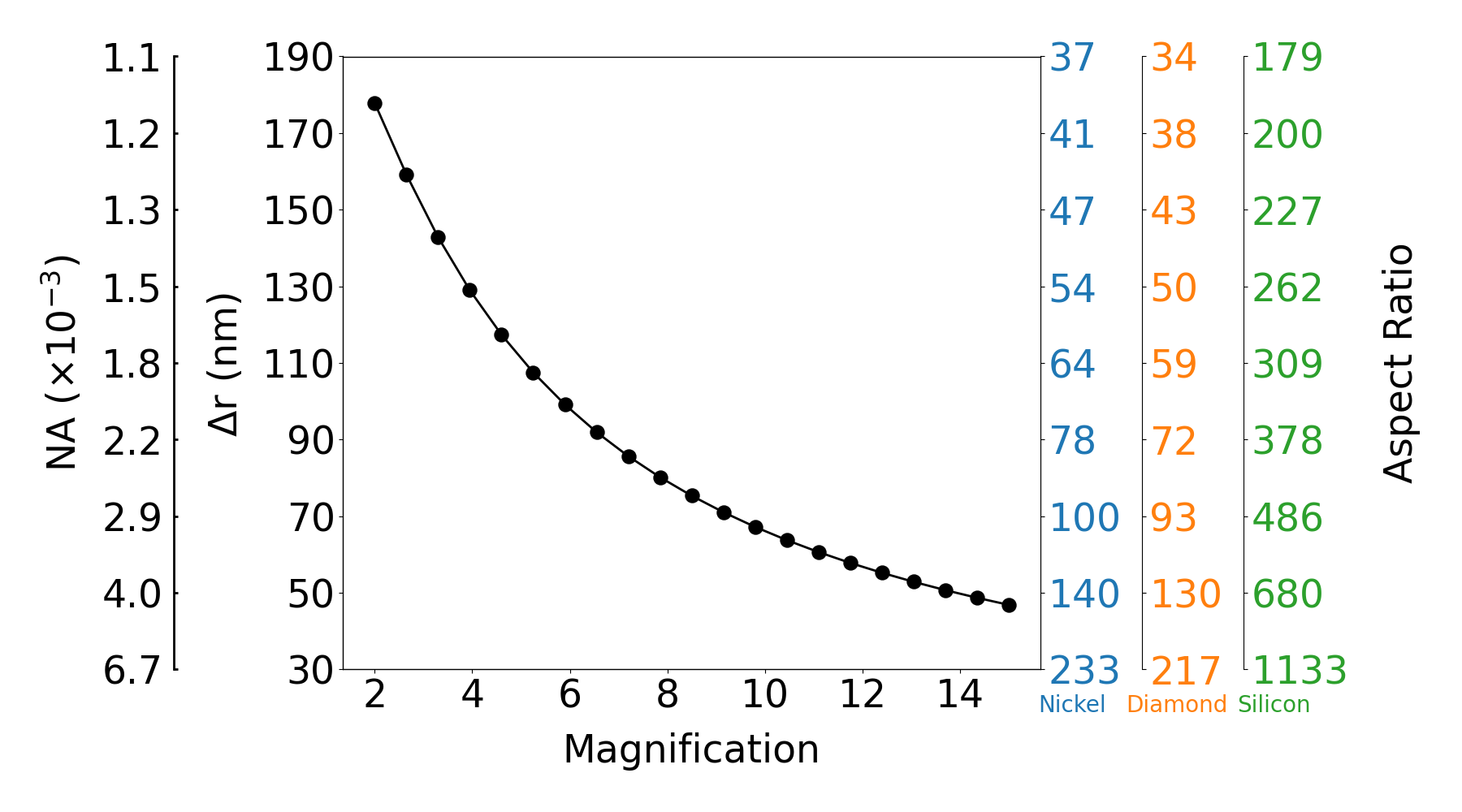}
    \caption{\small Manufacturing specifications for a single FZP objective in terms of outermost zone width $\Delta r$ as function of magnification for three materials. The corresponding  numerical aperture is shown. Moreover the required aspect rations of the outermost zone for three materials for maximum efficiency  is shown. The example relates to a fixed diameter $D = 4$ mm, wavelength $\lambda = $4 \AA{} and an experimental laboratory allowing $L = 8$ m.}
   \label{fig:FZP_manu}
\end{figure}

It appears from Eq.~\ref{Eq-FZP-T_pi}, \ref{FZP-f} and \ref{FZP-NA} that the critical performance parameter of the FZP is the \emph{aspect ratio} of the outermost zone, AR $ = \Delta T_{\pi}/ \Delta r$. Table~\ref{tab:opticals_constants_FZP} provides a list of materials that can be considered for the fabrication of neutron FZPs and key optical parameters. We have $\delta  \propto \lambda$. In Fig.~\ref{fig:FZP_manu}, as an example we show the manufacturing requirements to AR for the three materials chosen - and for a relevant setting of $D, \lambda$ and $L$.

\begin{table*}
    \centering
    \begin{tabular}{|c|ccc}
         Material &  $\delta (10^{-6})$ & $\mu$ (m$^{-1}$) & $\Delta T_{\pi} (\mu$m)\\
         Nickel &    24 &   260 & 8.34  \\
         Diamond & 29.8 & 97.9 & 6.71 \\
         Silicon & 5.28 &  13 & 37.8  \\
    \end{tabular}
    
    \caption{\small Neutron optical parameters for materials of relevance to manufacturing a neutron FZP - for a wavelength of  4 \AA. The linear attenuation coefficients represent the powder case.}
    \label{tab:opticals_constants_FZP}
\end{table*}

It can be seen from Fig.~\ref{fig:FZP_manu} that, to effectively match the incident divergence from the condenser, the FZP would require very high–aspect-ratio nanostructures, which are technically challenging to fabricate. The following strategies can be improved and developed to achieve these high aspect ratios: \cite{Chang2014}, \cite{Mohacsi16} and \cite{DeAndrade2021}. However, an FZP with  $\Delta r$ around 40~nm, roughly matches the NA$_{max}$ of a diamond CRL, and would still have 5 times smaller NA than the incoming divergence from the proposed condenser.


\subsection{A bank of FZP objectives}

Analogous to the CRL case discussed above, we propose addressing the mismatch between the numerical aperture of the FZP and the incident divergence of the condensed beam by employing a bank of FZPs. The considerations in relation to avoiding overlap of partial images, cf. Eq.~\ref{Eq-sampling}, apply here as well. 

\begin{figure}
    \centering
    \includegraphics[width=1\columnwidth]{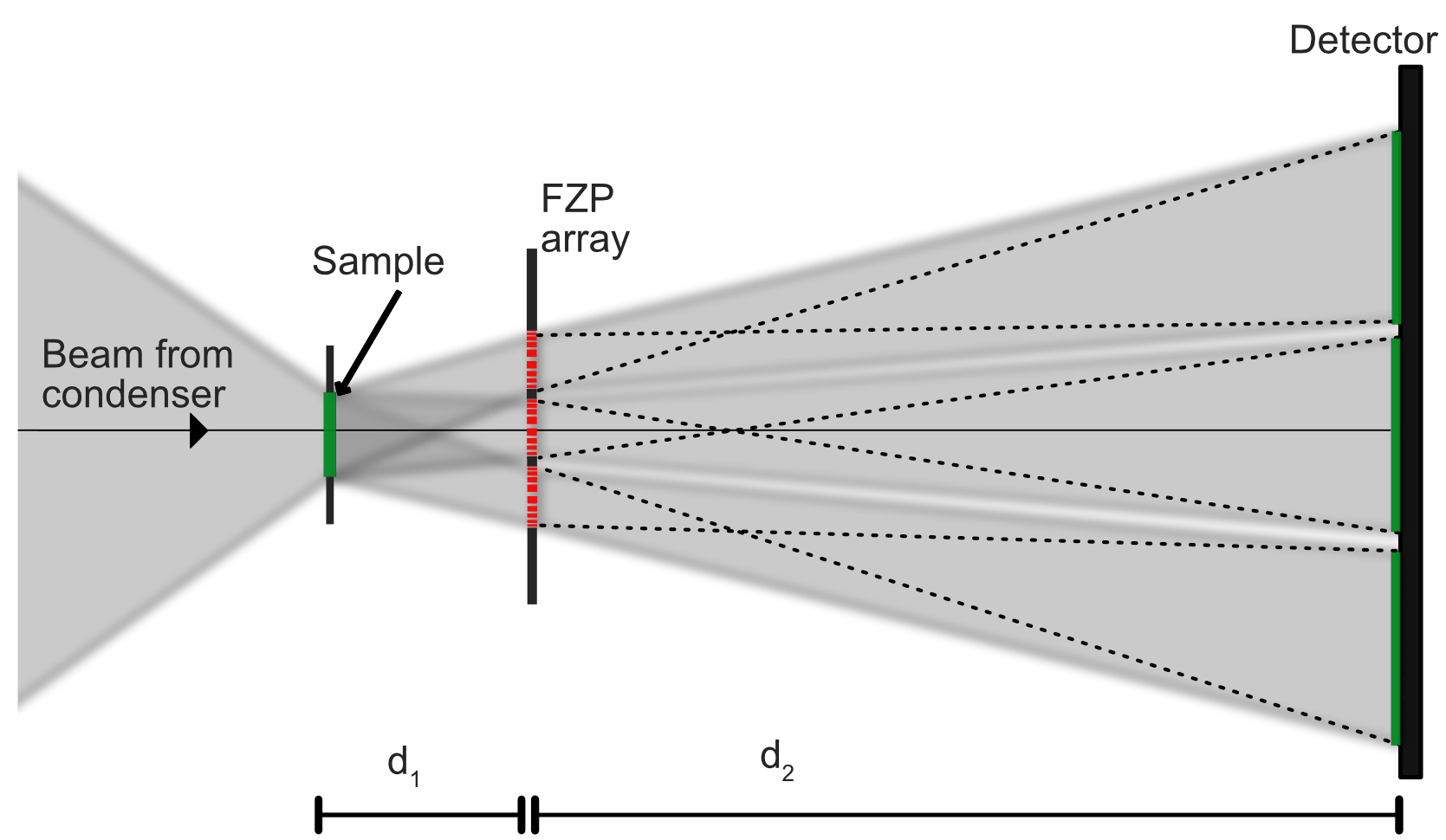}
    \caption{\small Illustration of the concept of a bank of objectives, where each objective is a Fresnel Zone Plate. The optical axes are in the case shown parallel.}
   \label{fig:FZP_bank}
\end{figure}

A solution based on the fabrication of FZPs in a regular grid with a pitch corresponding to the diameter $D$, on a single substrate, is illustrated in Fig.~\ref{fig:FZP_bank}. Such a design can be manufactured using existing lithography techniques. This approach provides a compact, easy-to-install solution and can be readily integrated into most existing instruments.
To prevent overlap between the images produced by individual FZPs, the FOV should be limited to the diameter of the FZP. With an NA of 0.002 (for $\Delta r$ = 100 nm) the number of individual FZPs required to match the full divergence is larger than for the corresponding CRL case, approaching 600. 
While demanding, we believe that the technical feasibility is primarily determined by that of an isolated FZP. 

It appears that the aberrations due to off-axis imaging will become worse for the outermost FZPs in the grid with the design shown in  Fig.~\ref{fig:FZP_bank}.  As mentioned we estimate that the aberration will still be tolerable, but this needs verification. To avoid off-axis aberrations, and ensure complete spatial and angular sampling, we consider fabrication of the bank of FZPs on a curved substrate - with a radius of curvature of $d_1$. Patterning an array of FZPs on a curved substrate has been demonstrated for visible light applications \cite{Moghimi2015, Low2022}. One concern for neutrons is that the high aspect ratio nanostructures when tilted on a curved substrate at some point will start to violate the Scalar diffraction approximation (\cite{Ali:20}). 

In conclusion, for monochromatic neutron beam operation, this type of objective bank is well-suited for inclusion in a micro-CT neutron microscope based on the Wolter-I condenser. The available space around the sample is adequate, it matches the field-of-view and
the magnification enables the combination of a micrometer-size spatial resolution with a magnification of 5 or more. In comparison to the CRL bank, the FZP solution on a flat surface is more elegant and technically easier to scale up. Simulations are required to learn if the off-axis aberrations are acceptable -  if not R\&D is required in FZP technology on curved surfaces. 

\section{An objective bank for large energy bandwidth imaging}
\label{sec-achromat}

CRLs and FZPs are both chromatic type optics. 
When changing the neutron energy the focal length and the magnification change. Superposing the signal from different energies leads to longitudinal and lateral (also known as transverse) chromatic aberration, respectively. The former manifests itself in a constant blur across the field-of-view while the latter blur increases with the distance to the optical axis. In the thin lens limit we have (with magnification unsigned)
\begin{eqnarray}
    \Delta d_2  & = (1 + \mathcal{M})^2 \;  \Delta f \hspace{3mm} \mathrm{Longitudinal} \; \mathrm{blurr} \\
    \Delta r_{det} & = r_{det} \dfrac{\mathcal{M}(1 + \mathcal{M})}{f} \;  \Delta f \hspace{3mm} \mathrm{Transverse} \; \mathrm{blurr} 
\end{eqnarray}
Here $\Delta r_{det}$ is the shift on the detector expressed in terms of the radius $r$ to the optical axis. (The different functional dependence of focal length $f$ on energy implies that the FZP exhibits only half the blur of the CRL.)

\begin{figure}
    \centering
    \includegraphics[width=0.7\columnwidth]{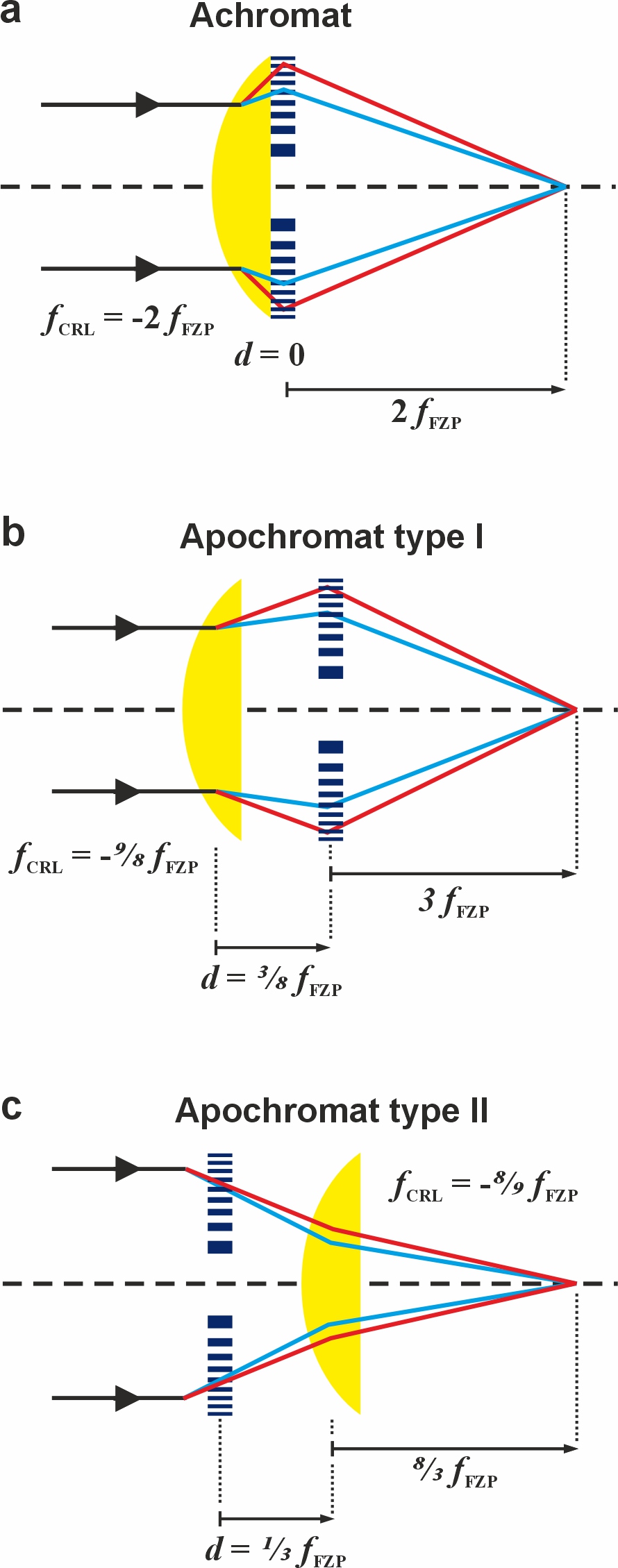}
    \caption{\small Sketch of the principle of a neutron achromat (a) and neutron apochromats of type I (b) and II (c) in a focusing geometry. Red and blue lines indicate rays with energies that are 10\% different. Relationships between key optical distances are indicated (see main text). Adapted from \cite{Sanli2023}}. 
   \label{fig_achro_apo}
\end{figure}

This chromatic aberration is  clearly a critical issue for high spatial resolution  imaging work, as it reduces the available flux by several orders of magnitude. In visual light optics, a combination of optics with different dispersion powers is used to enable larger bandwidth imaging. Similar optics have been introduced for X-rays: \cite{Skinner2001} and \cite{Wang2003} proposed achromats comprising a positive diffractive lens adjacent to a negative refractive lens.
In the thin lens limit, the focal length of the combined optics $f_{achr}$ is independent of energy to second order provided $f_{CRL} = - 2 f_{FZP}$. The resulting focal length is then $f_{achr} = 2 f_{FZP}$. This is illustrated in Fig.~\ref{fig_achro_apo} a). Likewise \cite{Skinner2004} and \cite{Chapman2021} discuss apochromats, where the two optics elements are separated by a distance $d$. By suitable tuning the combined "focal length" $f_{apo}$ becomes independent of energy to third order. 

The latter paper distinguishes two types of apochromats, referred to as Type I and Type II. In Type I the refractive lens is upstream of the diffractive lens - illustrated in Fig.~\ref{fig_achro_apo} b) - and in Type II the diffractive lens is followed by the refractive lens - Fig.~\ref{fig_achro_apo} c).  Key optical parameters, such as the distance from the exit of the combined optics to the focal point are summarized in the figure. Notably the entire optics becomes significantly longer than for the apochromat solutions and the FoV may also be impacted. The paper also extends the theory to thick lenses. 

Achromats and apochromats Type I were both experimentally demonstrated with X-rays by \cite{Kubec2022} and \cite{Sanli2023}. These optics were fabricated using electron beam lithography, electroplating and two-photon polymerization-induced lithography. High resolution scanning transmission X-ray microscopy (STXM) imaging has been demonstrated using both these lenses over a  wide bandwidth range, demonstrating the suitability of this class of optics for broadband, high-resolution imaging. 

In \cite{Poulsen2014} it is proposed to create broadband neutron optics in a similar way by combining a converging Fresnel zone plate  and a diverging CRL. In this section we first summarize these results. Then we explore the concept of a bank of neutron achromats.

\subsection{A neutron achromat or apochromat}
\label{sub-neutron_achromat}

In \cite{Poulsen2014} the optical properties of an achromat and a Type II apochromat are determined using ray transfer matrix analysis. Here the CRL is approximated by two parameters; the thin lens focal length and the thickness $NT$. By differentiating the combined focal length with respect to energy, \emph{analytical expressions for the thick lens case} are derived for the parameters for achromatic focusing (focal lengths, distance d etc). 

\begin{figure}
    \centering
    \includegraphics[width=1\columnwidth]{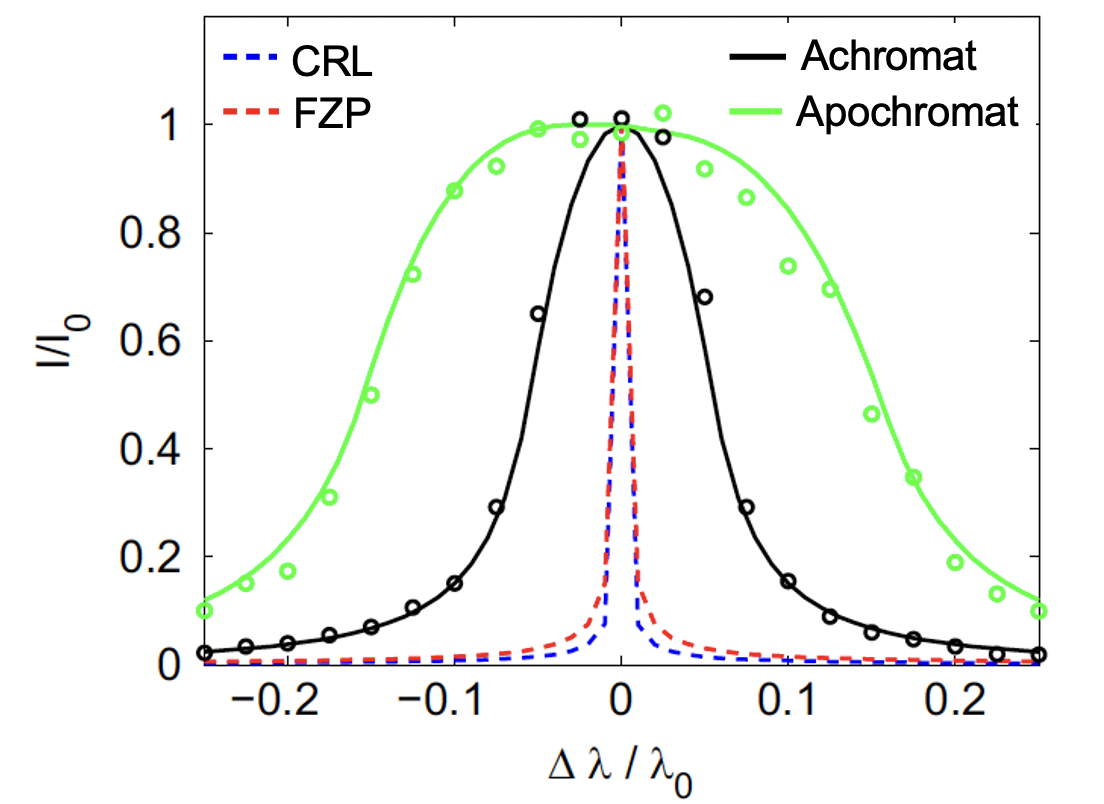}
    \caption{\small Chromatic aberration of a neutron achromat (black symbols) and a type II apochromat (green). Shown is the  variation in intensity in the image as a function of neutron wavelength. The dispersion of FZP and CRL optics (dashed lines) is compared to the two achromat solutions with the FZP and CRL connected or separated (full lines). Lines represent results from ray transfer matrix analysis, and circles are results from Fourier optics simulations. The figure is adapted from \cite{Poulsen2014}.}
   \label{fig:achromat_result2}
\end{figure}

Next this analysis is repeated for an imaging set-up. Notably, the parameters now depend on magnification. To illustrate this the relation between $f_{CRL}$ and $f_{FZP}$ for the achromat is
\begin{eqnarray}
    f_{FZP} = -\dfrac{f_{CRL}}{2} - \dfrac{NT}{6}\dfrac{2\mathcal{M}-5}{\mathcal{M}-1}, \label{eq-focal_achro} 
\end{eqnarray}
which in the thin lens limit corresponds to the expression in Fig.~\ref{fig_achro_apo} a).

The paper proceeds with detailed simulations of a $\lambda = 6$ \AA{} set up with a fixed 5~mm diameter FZP, a focal length of 1~m and a magnification of 3. An example of the results is reproduced in Fig.~\ref{fig:achromat_result2}. The energy bandwidth, as defined by the FWHM, is 12 \% and 31 \% in the two cases. This work also documented that the efficiencies of the achromat/apochromat were almost as good as the FZP on its own (37 \% efficiency for both achromat and apochromat.)   This entire work was further substantiated with Fourier optics simulations, validating the geometrical optics work.  

In order to minimize the wavelength-dependent distortion due to gravitation, a neutron prism (wedge) could be used as both these effects grow proportional to $\lambda^2$, see \cite{Hammouda2007}. For an implementation at a ToF-based imaging instrument such as ODIN, the gravitational effects may  also be corrected for during the data analysis.

\subsection{A bank of neutron achromats or apochromats}
\label{sub-design}

Based on the results of sections \ref{sec-CRLobjective} and 
\ref{sub-neutron_achromat} we can now discuss the prospect of an objective comprising a bank of neutron achromats or apochromats. 

\begin{itemize}

\item \emph{Spatial and angular coverage.}   FZPs and CRLs can be made to match in terms of NA, FOV and focal lengths. The achromat and apochromat solutions can provide a full coverage of the angular range defined by the divergence of the condensed beam. The small angular gaps in the coverage between the individual units correspond to finite angular step sizes, which is not an issue for the reconstruction algorithm. The field of view of each unit will be governed by the CRL and will match the specified 4 $\times$ 4 mm$^2$. If possible, to avoid cross talks between units the incident beam should be confined to 4 $\times$ 4 mm$^2$ by a slit. 

\item \emph{Efficiency and energy band.} As mentioned the CRLs will effectively be transparent lenses, implying that the efficiency is governed by the zone plates. Theoretically the overall efficiency can be better than 30 \%. The theoretical bandwidth for achromats and apochromats are displayed in Fig.~\ref{fig:achromat_result2}.

\item \emph{Spatial resolution.}  The diffraction limit and known aberrations in the FZP are all below  2 $\mu$m. This specification is also easily achievable in terms of vibration control and temperature stability. Hence, the limitation on spatial resolution will be precision in manufacturing and assembly. Assuming a flat surface design is used, the key performance parameter for the FZP is the aspect ratio of the outermost zone, cf. Fig.~\ref{fig:FZP_manu}. The divergent CRL is the inverse of the design shown in Fig.~\ref{fig:CRL_objective__bank} - it can be realized e.g. by free standing but identical parabolic shaped balls made of diamond or MgF$_2$. These balls are then organized in space by means of mechanical guides. The design is beyond the scope of this article, but we assume that a manufacturing accuracy of a few micrometer is within reach. 

\item \emph{Detector technology.}  The long focal lengths of the achromat/apochromats implies that the magnification  will be below 10. A CMOS camera coupled to a scintillator screen is well adapted to this case in terms of pixel size. The neutron microscope will require a massive coverage by detectors, defined by say 400 simultaneous projections each aiming at resolving a 1000x1000 area. Adapting the focal lengths for the individual projections all data may be acquired on one large planar detector with say 1 Gpixels. Notably, for X-ray imaging scintillator based cameras with 150-600 Mpixel and 1-4 $\mu$m pixel size are now in routine use \cite{Kameshima2022, Gellert2025}. A similar technological development within neutron imaging detectors can be expected in the coming years.

As an alternative, one may aim to increase the magnification such that single-neutron-counting detectors with larger pixel sizes become viable. This will improve the S/N ratio. However, the cost of hundreds of detectors may be prohibitive.

\item \emph{Choice of broadband optics.} 
The three alternative set-ups illustrated in Fig.~\ref{fig_achro_apo} have different merits. The achromat solution is superior in relation to neutron magnification, cf. the discussion on detector above. The apochromats provide a broader energy bandwidth. 

\item \emph{Versatility.}  The average energy is fixed with the achromat solution. For a thick lens so is the magnification, cf. Eq.~\ref{eq-focal_achro}, unless a tuning opportunity is added. In the apochromat solution, the added degree of freedom does in principle allow for tuning the energy but the complication of adapting focal lengths may disfavor this solution.  We propose instead to think of the entire objective bank as a fixed optics with no internal degrees of freedom. This optics will be  confined in space (approximately $ 0.1 \times 0.1 \times 2$ m$^3$) and weight. Hence, it can be inserted and removed at will - e.g. enabling multiscale imaging of large objects. In principle several such optics could be manufactured to ensure versatility in terms of energy and magnification.

\end{itemize}

\section{Discussion}

A detailed optimization study of the design of the neutron microscope is beyond the scope of this article. We have listed the perceived key limitations and provided a discussion of the feasibility. As a follow up we are currently working on the conceptual design of an instrument for the European Spallation Source (ESS), which would be optimally suited to take advantage of neutron microscopy and set-ups that have been discussed here. The instrument is intended to be relatively short and to provide a homogeneous cold neutron beam of high divergence and flux. The relaxed wavelength resolution would on the one hand be sufficient to enable gravity correction in data post-processing, and on the other hand to support efficient implementation of advanced neutron contrast modalities such as dark-field contrast, inelastic scattering contrast and depolarisation contrast imaging, which have a high potential to profit from a neutron microscopy set-up paving the way for pioneering high resolution investigations providing unique material science information not accessible today.  

The current work plan for the neutron microscope includes
\begin{itemize}
\item A test of a prototype condenser with a design that is slightly different from the one used in the simulations in section 2.1.
\item Fabrication of a nickel FZP array on a flat substrate as shown in Fig.~\ref{fig:FZP_bank} using electron beam lithography and electroplating techniques as a first step and proof-of-principle demonstration of multi-plexed neutron imaging. 
\item A full scale simulation of a complete microscope design.
\end{itemize}

\subsection{Condenser}

The Wolter-I optic design considered here for the neutron condenser was optimized for telescopes, i.e. for an extremely low divergent beam.  An elliptic-hyperbolic combination of Wolter optics, would be better suited for divergent beams \cite{Liu2012}. Moreover, specifically in relation to ODIN, it is obvious that a better match can be made between the exit of the neutron guide and the dimensions of the optic, e.g. by replacing the elliptical part by a straight section. The rationale for the choice of model system in section \ref{s:rt} made is that all parameters are copied from an existing instrument. This gives confidence in the conclusion that a gain in neutron flux of up to a factor 100 is feasible.   

We emphasize that the neutron condenser is relevant without the objective.  When applied in front of a pinhole camera type instrument, it enables increasing the beam divergence homogeneously and thus achieve a larger field of view despite a relatively low incoming divergence, as found typically in neutron imaging instruments at the end of a neutron guide. It can be used as a means of efficient pinhole to pinhole beam transport, for efficient scanning type mapping experiments, for high resolution measurements of thin samples being basically in contact with the detector, for medium resolution high intensity imaging and e.g. for high resolution neutron capture imaging.


\subsection{Objective} 

A key advantage of any microscope is that it is associated with a Fourier plane, the Back Focal Plane (BFP).  In electron, light and X-ray microscopy, operations in the BFP  have led to a range of modalities for improving contrast or angular resolution.  For the multiplexed objective such operations can be provided in each achromat unit. These may be the same but can also be set to be different, thereby allowing for simultaneous multimodal imaging.  

In terms of alternatives to the proposed optics, we emphasize that reflective optics being achromatic are ideal. However, also for reflective optics a multiplexed objective design may be relevant to improve angular resolution.  We note that a full-field X-ray microscope based on Kirkpatrick–Baez mirror optics has shown sub-micrometer resolution over a broad energy range \cite{Matsuyama2019}. However, such solutions require the use of a thick substrate and we are not familiar with solutions that allow multiplexing such optics. 

We also mention that an alternative solution to overcoming the issue of achromaticity is to exploit the fact that neutrons have finite velocities and change configuration of the objective during the duration of the pulse. This is discussed in \cite{Poulsen2014}.
 
Finally, we emphasize that the objective bank is relevant without the condenser and in fact may be used for SANS imaging with a highly collimated incident beam.  In this case the specifications for NA can be relaxed in order to increase the Q range.

\subsection{Implementation in neutron instruments}

The condenser is relatively confined in length and the objective bank is confined in terms of angular coverage to 3 deg. Hence, both types of optics may be tested and find use at existing neutron instruments. 

For the dedicated neutron microscope two prime concerns will be the length of the lab and the compliance with the neutron guide - in order to optimize flux density on sample. 

\subsection{Data analysis}

The high divergence of the beam on the sample implies that the depth of focus will often be smaller than the thickness of the sample. This implies that the classical cone-beam geometry ansatz in tomography does not apply. Moreover, some of the projections provided by the objective bank will be out-of-plane by up to 1.5 degrees. Hence, tomographic reconstruction will require a new algorithm - this however is straightforward using a linear algebraic approach and an optimization based on the resemblance of experimental data and simulated data from a forward model of a digital sample \cite{Hansen2021}.  \emph{A priori} information about the sample structure may be included e.g. by adding a regularizing term.

The Crowther criterion states that to provide a 3D tomographic reconstruction of an object of diameter $D$ with a spatial resolution of $\Delta x $, the angular step size $\Delta \omega$
during a 180 deg rotation must fulfill
\begin{eqnarray}
 \Delta \omega \le \dfrac{2 \; \Delta x}{D} \; \; \approx \; \; \dfrac{2}{N_{voxel}}
\end{eqnarray}
 where $N_{voxel}^3$ is the number of voxels in the reconstructed volume. In the achromat solution presented in section \ref{sec-achromat} projections are acquired with an angular width
given by the NA of the FZP. Notably 4 mrad corresponds to $N_{voxel} = 500$, which is well matched to many science cases. In contrast, if projections implied integration over the full divergence of 3 deg, $N_{voxel} = 35$, which is often too small. 
For this reason we argue that the multiplexed objective at times may be favorable even if it becomes technically feasible to make a nested mirror type objective. 

The fact that the beam is focused within the sample also opens up for other types of scans including depth scanning and local tomography.

\subsection{Science case}

Access to micrometer resolution with high neutron flux based on the proposed optics paves the way to new material science, addressing key properties and processes only accessible by neutrons. These range from industrial to natural and cultural heritage materials and cover for instance bulk metals and ceramics, which X-rays cannot penetrate, but also soft and biological structures that high intensity X-ray beams would destroy in the course of a measurement.    

\begin{itemize}
    \item \emph{Imaging with significantly higher neutron flux.} Operando neutron imaging provides key observations in a range of energy conversion and storage technologies addressing mechanisms underlying performance, degradation, life-time etc. for batteries and fuel and electrolysis cells. The significantly increased neutron flux enables real-time operando analyses of the devices exploiting the unique neutron contrast for hydrogeneous materials, low-atomic number ions in new materials for rechargeable batteries etc. Similar cases exist for additive manufacturing, flow in porous media etc.
    \item \emph{Imaging with polarized neutrons.} Achieving high spatial resolution with polarised neutron imaging techniques that require a polarisation analyser is a challenge with a pinhole geometry. The analyser and necessary magnetic field decoupling require a significant distance between sample and detector. Due to the limited available flux this typically limits spatial resolution to a few hundred micrometers. Microscope optics can overcome this severe limitation enabling insight into micrometer range magnetic order and transitions, and current distributions.
    \item \emph{Diffraction-based imaging.} Here an objective-based solution has the fundamental advantage that spatial and angular degrees of freedom decouple. By scanning angular degrees of freedom, a microscope allows for determining the center-of-mass orientation and/or strain components for each voxel in the sample with high accuracy.  
\end{itemize}

\section{Conclusion}

The large divergence and large source size inherent to neutron sources has been a major limitation in the development of full-field neutron microscopy. In order to fulfill the simultaneous requirements to spatial resolution, working distance, efficiency and broad band operation, we have presented a design involving a combination of a nested super-mirrors condenser and an objective bank comprising neutron achromats. We argue that this solution is valid in terms of optics principles and we have discussed challenges and solutions for implementation. For tomography a key advantage is that hundreds of projections are acquired simultaneously with an angular integration that match reconstructions of $1000^3$ voxels. The optics concepts are also valid for diffraction based imaging.  

\section{Acknowledgment}

We thank Robin Woracek, Jochen Stahn, Sina Maria Baier-Stegmaier and Kurt Clausen for scientific discussions. We acknowledge PSI and the 3D  imaging centre at DTU for beamtime. 
HFP, EBK and LTK acknowledge financial support from the Danish ESS lighthouse on hard materials in 3D, SOLID. LTK and HFP acknowledge funding from the Novo Nordisk Foundation grant no. NNF25OC0100016. CHEXS acknowledges financial support from the Innovation Fund Denmark. This work also received funding from a PSI research grant 2021.
\bibliographystyle{plain}
\bibliography{references}
\end{document}